\documentclass[twocolumn]{aastex7}

\usepackage{graphicx} 
\usepackage{amsmath}

\newcommand{\be}{\begin{equation}}
\newcommand{\ee}{\end{equation}}
\newcommand{\bel}[1]{\begin{equation}\label{#1}}
\newcommand{\ba}{\begin{eqnarray}}
\newcommand{\ea}{\end{eqnarray}}
\newcommand{\bal}[1]{\begin{eqnarray}\label{#1}}

\newcommand{\Msun}{M$_\odot$}

\begin{document}

\title{What is the most massive gravitational-wave source?}
\author[0000-0002-6134-8946]{Ilya Mandel}
\affiliation{School of Physics and Astronomy, Monash University, Clayton VIC 3800, Australia}
\affiliation{OzGrav: The ARC Centre of Excellence for Gravitational Wave Discovery, Australia}
\email{ilya.mandel@monash.edu}

\begin{abstract}

In the presence of significant measurement uncertainties, the events which appear to be the most extreme are very likely to be those exhibiting the greatest statistical fluctuations.  It is therefore particularly important to exercise care when interpreting such events and to use the entire observed population for context.  Here, I attempt to pedagogically illustrate this using the example of the most massive binary black hole so far detected in gravitational-wave data, GW231123.  I argue that its total mass may be significantly lower than $238^{+28}_{-49}$ \Msun\ as reported by \citet{GW231123}.  The maximum total binary black hole mass from an analysis of the entire detected population is below $170$ \Msun\ if the same priors that are used for individual event analyses in the GWTC catalogs, including for the analysis of GW231123, are applied to the population as a whole.  However, this value is very sensitive to assumptions about the population distribution.

\end{abstract}

\section{Introduction}

\citet{GW231123} announced the discovery of GW231123.  A Bayesian analysis of the gravitational-wave signal from the merger of this binary black hole yielded a total mass of $238^{+28}_{-49}$ \Msun, where the value denotes the median of the posterior and the error bars denote the edges of the 90\% credible interval.  This made it the most massive binary black hole merger observed to date by the LIGO-Virgo-KAGRA collaborations (LVK), surpassing any of the objects released in previous gravitational-wave transient catalogs \citep{GWTC2, GWTC3}.  In fact, it was pre-reported before the full release of the GWTC4 catalog \citep{GWTC4}, to which it belonged, as an exceptional event.

Here, I attempt to pedagogically discuss why the interpretation of extreme events warrants some additional care.   

Bayesian posterior probability distributions are meant to guarantee that, if the likelihood function is perfectly computed and if the true parameters of events are fairly drawn from the assumed prior distribution (we will return to these non-trivial assumptions later), then, taken as a whole, over the observed population, the credible interval are unbiased.  Unbiased posteriors mean that one should expect that the $X$\% credible interval contains the true parameter for $X$\% of observed events -- something often checked with p-p plots when testing parameter estimation pipelines \citep[e.g.,][]{Sidery:2013}.  A Bayesian analyst also expects that the quantile of the truth in the posterior is uniformly distributed on $[0,1]$ across all events.  

However, it should not be expected that an individual event posterior is centered on the true value.  If it were, we would know the true parameter values of individual events just by looking at the posterior medians!  

In fact, we should expect that posteriors of extreme events are particularly shifted in the direction that makes them extreme.  For example, if by luck we happened to observe 150 events that all had identical true masses and identical measurement uncertainties $\sigma$, we should not be surprised if the highest-mass event has a posterior mean that is $\sim 2.5 \sigma$ above the truth.  This point was eloquently made by \citet{Fishbach:2020} in their analysis of the maximum primary black hole mass from the first 10 gravitational-wave detections.

Thus, the mass estimate of GW231123 is prima facie likely to be biased upward because it is the event whose mass posterior favors the highest masses.  This then raises the question of what is actually the most massive binary black hole in the population based on existing gravitational-wave data.  In section 2, I address this question by considering the priors used by the LVK for data analysis; in section 3, I discuss how modifying those priors may impact the answer; section 4 contains brief conclusions.

\section{Inferring the maximum binary black hole mass using LVK priors}

As mentioned previously, Bayesian posteriors should be unbiased for the population if the likelihoods are correctly computed and the true parameters are really drawn from the assumed priors.  These are highly non-trivial assumptions.  

Correctly computing the likelihood for gravitational-wave events means perfectly knowing the waveforms, i.e., the translations from model parameters such as masses and spins to the gravitational-wave signal from a merging binary.  It is clear that these are not perfectly known.  \citet{GW231123} discuss large waveform systematics: shifts in parameter estimates due to differences in the waveform models used for analysis.  Moreover, previous experience shows that the choice of which physics to include in the models can impact the results; for example, \cite{Miller:2024} concluded that GW190521 was a precessing binary from an analysis with waveforms that allowed for precession but not eccentricity, while \citet{RomeroShaw:2020GW190521} found that it was an eccentric binary on the basis of an analysis with waveforms that allowed for eccentricity but not precession.  Furthermore, the likelihood function must correctly include the translation between the actual observable signal and the strain waveform, known as calibration \citep[e.g.,][]{Vitale:2021,GWTC4:data} and must accurately model detector noise, which is likely non-stationary, non-Gaussian, and may include artefacts such as glitches \citep[e.g.,][]{Roever:2011,Payne:2022}.  However, for the purpose of this discussion, I shall assume that the likelihood function is correct.

The mass priors used in the LVK data analysis are flat in the component masses.  Technically, these are detector-frame component masses (i.e., multiplied by $(1+z)$, where $z$ is the source redshift that is simultaneously estimated), and there are additional cuts, such as those placed on the mass ratio because waveform models have limitations in their accuracy at extreme mass ratios.  Moreover, the choice of priors on parameters other than masses can impact mass measurements \citep[e.g.,][]{Mandel:2025}.  However, to keep this discussion simple and pedagogical, I will only consider the total source-frame mass $M$.  If component masses are uniform over the range $[0, M_{c,max}]$, then the total mass prior follows the distribution
\bel{pM}
p(M) = \frac{4}{M_{max}^2} 
\begin{cases}
      M & \text{if}\ 0 \leq M \leq M_{max}/2 \\
      M_{max}-M & \text{if}\ M_{max}/2 \leq M \leq M_{max}\\
      0 & \text{otherwise}
    \end{cases}
\ee
where $M_{max} = 2 M_{c,max}$.

Suppose for now that the shape of this prior is accurate, i.e., the true total masses of detectable gravitational-wave events follow this distribution.  We can then attempt to learn the true value of $M_{max}$, the maximum total mass in the population of detectable binary black holes, from the full set of observations to date.  This is a very simple example of hierarchical modelling.  

As pointed out by \citet{EssickFishbach:2023,Toubiana:2025}, selection effects can impact population inference even on the {\it detectable} population.  Consider, for example, a fisherman who is fishing in a lake where only fish with a mass above 1 kg may be kept, while all lighter fish must be released.  The fisherman has a scale and follows this rule, weighing all caught fish, keeping those heavier than 1 kg and writing down their measured masses, while releasing lighter ones without taking notes.  However, the scale is not perfect, with a measurement uncertainty of 100 g.  If fish with measured masses of 1.043, 1.016, 1.015, 1.009, and 1.094 kg are recorded, we might conclude that catchable (``detectable’') fish have true masses that are distributed between 1 and 1.1 kg, perhaps with some preference for the lower end of this range, if we know nothing about selection effects.  On the other hand, knowing about the selection effects, we might conclude that the catchable fish have a mass distribution centered on a value closer to 0.9 kg (in fact, the values were drawn from a delta function at 900 g after imposing a Gaussian measurement uncertainty of 100 g and selection effects).

This example is somewhat extreme in imposing a hard cut.  In practice, if the search sensitivity is a relatively flat function of the total mass in the range of interest, as suggested, for example, by figure 6 of \citet{Belczynski:2014VMS}, we can eschew selection effects when considering a hierarchical model of the total mass distribution of the population of {\it detectable} merging binary black holes.  This is the approach we take here.  However, a more careful treatment should include the impact of selection effects as in \citet{Toubiana:2025} (see, e.g., \citealt{Mandel:2018select} for a discussion of hierarchical modelling in the presence of both selection effects and measurement uncertainties).

Consider a hierarchical model of the total mass distribution of the population of detectable merging binary black holes.   Assuming a flat prior on the population-level parameter $M_{max}$, its posterior is given by
\be
p(M_{max} | \{data\}) \propto \Pi_{i=1}^N p(d_i | M_{max}),
\ee 
where the full data set consists of $N$ events $d_1$...$d_N$.  The likelihood for a given event $d_i$ given the population model parametrized by $M_{max}$ is
\bel{integral}
p(d_i | M_{max}) = \int  p(d_i | M) p(M|M_{max}) dM.
\ee
Here, $p(M|M_{max})$ is given by Eq.~(\ref{pM}) while the likelihood $p(d_i | M)$ can be obtained, up to normalisation, by dividing the LVK posterior by the priors $\pi_{LVK}(M)$ used in the original analyses.  In practice, the integral in Eq.~(\ref{integral}) can therefore be estimated as a Monte Carlo sum over the available LVK posterior samples:
\be
p(d_i | M_{max}) \propto \frac{1}{N_s} \sum_{k=1}^{N_s} \frac{p(M_k | M_{max})}{\pi_{LVK}(M_k)}.
\ee
where $N_s$ is the number of posterior samples $\{M_k\}$ for event $i$.
I will assume that the LVK used priors that are flat over a sufficiently broad range that the total mass prior is $\pi_{LVK}(M) \propto M$ in the range of interest; this is broadly consistent with the released prior samples.  I implicitly marginalize over other parameters.  As discussed below, the results are insensitive to these simplifications.  I use all events with source-frame total mass posteriors released to date\footnote{\url{https://gwosc.org/eventapi/html/GWTC/}} with false alarm rates less than 1 per year, using the LVK-default ``mixed''  samples from analyses with different gravitational-waveform families.

\begin{figure}
    \centering
    \includegraphics[width=\linewidth]{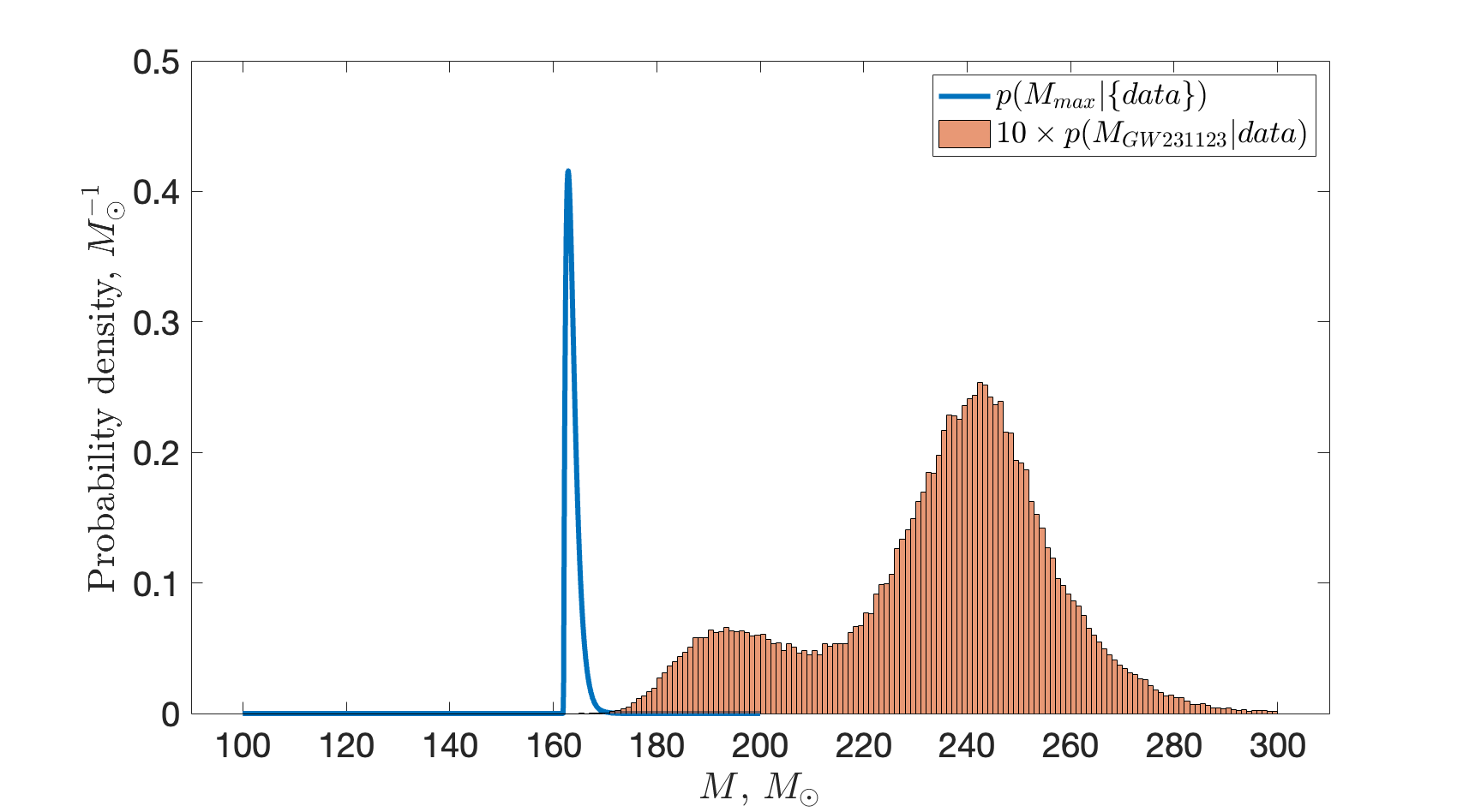}
    \caption{The posterior probability distribution on the maximum total mass $M_{max}$ given the LVK data from GWTC1 through GWTC4 (blue) and the histogram of the LVK ``mixed'' posterior on the source-frame total mass of GW231123 from a combination of analyses with different gravitational waveforms (orange, scaled up by a factor of 10 for improved visibility).}
    \label{fig:Mmax}
\end{figure}

The results of this analysis are shown in Figure \ref{fig:Mmax}.  The posterior on $M_{max}$ has a 90\% credible interval of $167^{+4}_{-1}$ \Msun.  In other words, if the component masses really followed flat priors, our best guess for the maximum total mass of a detectable binary black hole based on the full available LVK data set would be just $167$ \Msun.  This, of course, is very significantly lower than the claimed mass of GW231123.  

In fact, as seen in Figure \ref{fig:Mmax}, the posterior on $M_{max}$ rails against the lowest total-mass posterior samples for GW231123.   Of course, the maximum population mass cannot be lower than the minimal range of posterior support of all events, i.e., is limited from below by the largest value of the minimum of the posterior support among all events, in this case GW231123.  The priors $\pi_{LVK}$ used by the LVK and the selection effects do not vary significantly over the narrow range of the posterior on $M_{max}$, justifying our simplifying assumptions in this study.  The large systematic differences in the posteriors on the parameters GW231123 obtained with different waveform families translate into a shift in the best-fit value of $M_{max}$ by up to $\sim 20$ \Msun.  The narrow posterior on $M_{max}$ and its sensitivity to the GW231123 measurement is consistent with the view that, if we assume that the binary black hole masses follow a relatively compact underlying population distribution (i.e., one without heavy tails), then we are pushed toward the conclusion that GW231123 was a large statistical outlier.

\section{Adjusting the population priors}

\begin{figure}
    \centering
    \includegraphics[width=\linewidth]{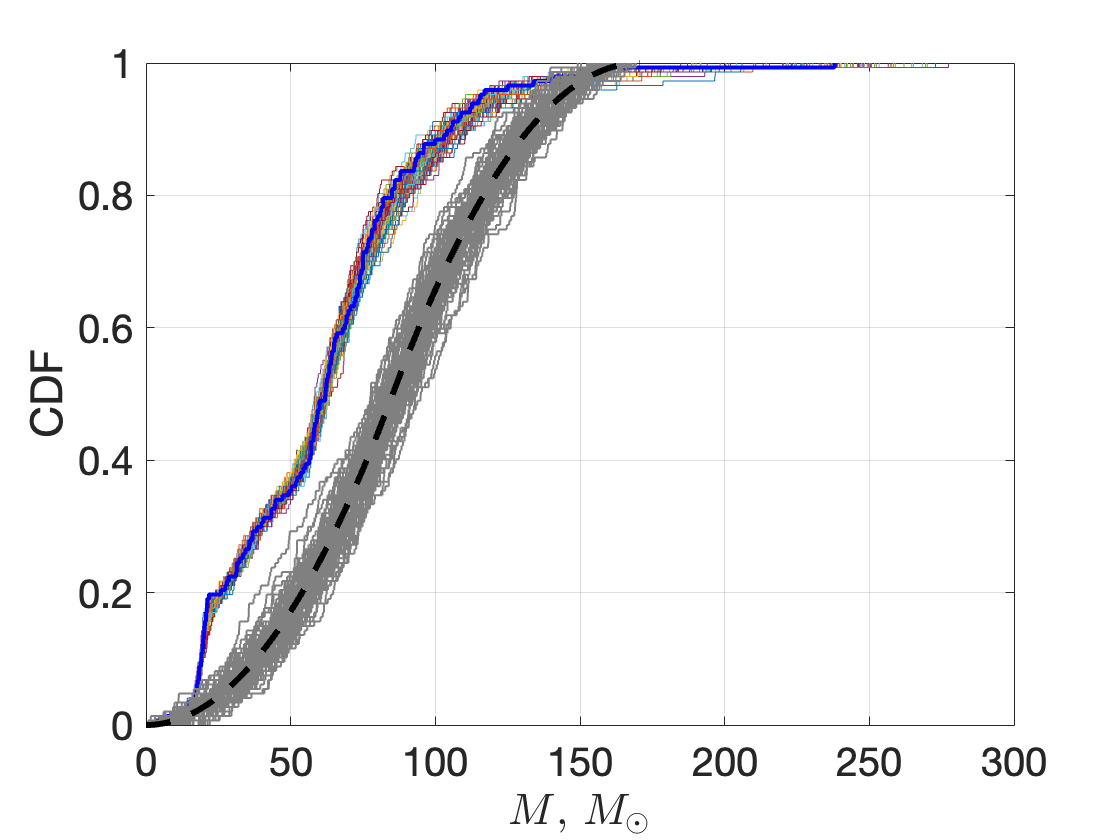}
    \caption{The cumulative distribution of total masses (thin colored background lines), the cumulative distribution of median total masses (blue line), the analytical cumulative distribution function from the flat-component-mass prior using $M_{max}=167$ \Msun (dashed black line) and cumulative distribution functions of draws from this model containing the number of observed events in the GWTC catalogs (grey lines).}
    \label{fig:CDF}
\end{figure}

As mentioned above, the Bayesian posteriors on individual events are unbiased only if the actual true parameters of events are drawn from the priors used in the Bayesian analyses.  However, this does not appear to be the case here.  Figure \ref{fig:CDF} shows 100 super-imposed cumulative distribution functions obtained by randomly sampling one total mass value from the posterior of each event as computed by the LVK with thin colored lines.  The cumulative distribution function of the medians of the individual event posteriors from the LVK analysis is shown with a blue line (this is provided for illustration only).  The dashed black line shows the cumulative distribution function constructed from the probability distribution function of Equation \ref{pM} using the best-fit $M_{max}=167$ \Msun, while 100 grey lines show cumulative distribution functions constructed by drawing as many observed events from this model as are present in the GWTC catalogs.   These model cumulative distributions, which account for the statistical fluctuations due to limited event counts, are clearly well outside the plausible range of empirical cumulative distribution functions that account for measurement uncertainties; the theoretical model would be readily ruled out by the Kolmogorov-Smirnov test.  This suggests that perhaps we should not take too seriously inference using this ill-fitting population model; by the same token, however, this also indicates why we should be weary of the analysis of GW231123 which used flat component mass priors that yielded Equation \ref{pM}.

Figure \ref{fig:CDF} also illustrates why the maximum mass is effectively set by the smallest allowed mass for GW231123, i.e., the smallest available LVK posterior samples for that event.  The medians of LVK posteriors for the eight events with the largest median values are 116 \Msun, 117 \Msun, 125  \Msun, 134  \Msun, 140 \Msun,  153  \Msun, 153 \Msun,  238 \Msun. GW231123 is a substantial outlier in terms of its total mass posterior, though see \citet{GW231123} for discussion of its standing relative to multi-dimensional population models.  A data analyst is forced to make a choice.  Is GW231123 a statistical fluctuation, where the measurement excursion happens to be more extreme than one might normally anticipate given the $\sim 150$ available detections?  Or is it genuinely physically exceptional, representing an entirely different population (\citealt{GW231123} also found that its black holes may have exceptional spin properties, though the very few in-band gravitational wave cycles make inference challenging)?    

\citet{GWTC4:pop} consider this question and conclude that ``while GW231123 lies in the tail of the distribution, its total mass is consistent with the inferred mass spectrum''.   Below, I consider the maximum binary black hole mass if, indeed, GW231123 belongs to the same population as other massive detected binary black holes. 

The plan is to model the overall distribution first, then re-consider the values of individual events in light of the revised priors produced by such a fit.  

As a simple analogy, consider asking the members of your department to measure and report the heights of their desks.  You exhort all of them to be good Bayesians and to report the posteriors on their desk heights after assuming flat priors between 0 and 200 cm.  When the reports come in, you note that most of them look like $79 \pm 1$ cm, $81 \pm 0.7$ cm, $83 \pm 2$ cm... Almost all are consistent with desks of 80 cm in height.  There is, however, one report of $140 \pm 30$ cm.  Admittedly, this is not a very precise measurement.  And it might correspond to some very special outlier, such as a standing desk.  But, given that the vast majority of desk heights are clearly consistent with 80 cm, it may be time to revise your prior to include a strong peak around 80 cm, albeit perhaps with some remaining support elsewhere.  When re-analyzed with this updated prior, you may find that the desk height previously claimed to be $140 \pm 30$ cm is now, say, only $85 \pm 10$ cm.  Note that there is no issue with double-counting the same data for prior and likelihood here; the prior for each measurement can be constructed using only the other measurements excepting the one being analyzed. This update to the measurement of an individual event based on the population properties is a natural outcome of hierarchical population modelling  \citep[e.g.,][]{Fishbach:2020,MooreGerosa:2021}\footnote{The actual approach taken by the LVK is described in \url{https://dcc.ligo.org/LIGO-T2100301/public}, \url{https://github.com/farr/Reweighting}.}.  Statisticians sometimes refer to the improved estimates on the parameters of individual events afforded by considering the full population as ``shrinkage'' \citep[e.g.,][]{StenningVanDyk:2018,Shariff:2016}.

This is one reason why it can be tricky to evaluate exceptional events before the full catalog is released: without a full population analysis, it is not possible to check whether the priors used for the analysis are consistent with the data or to update them if they are not.

Here, I will not attempt to construct a full model of the total mass distribution of detected events \citep[see, e.g.,][for such population models of other parameters]{GWTC4:pop}.  In any case, it is reasonable to assume that events in very different mass ranges may follow different distributions (e.g., because the binary black holes are formed through different evolutionary channels, \citealt{Mapelli:2021, MandelFarmer:2018}).  We can, however, attempt to model just the heaviest 5\% of events.  Given the uncertain measurements, I do not know which events these are and should marginalise over that uncertainty, but for simplicity, I will just consider the 8 events with median LVK posterior total masses $>115$ \Msun, which, of course, include GW231123.  The decision to select events based on the median LVK posterior masses is arbitrary and cannot be guaranteed to select the heaviest binaries; however, the conclusions are not sensitive to the details of this procedure and GW231123 is unambiguously the most massive observed event.

The tails of the distribution are always challenging to model and deciding on the shape of the revised prior is up to the analyst.  For illustration purposes, I will first assume that these 8 events follow a distribution that is flat in total mass between 115 \Msun and $M_{max}^{flat}$.   I can now create a new hierarchical model for the 8 most massive observed events, assuming their underlying total mass distribution is flat between 115 \Msun and $M_{max}^{flat}$.  Assuming a flat prior on $M_{max}^{flat}$, its credible interval is $190_{-12}^{+9}$ \Msun.  Thus, a population model based on the highest-mass events indicates stronger support for a more massive tail in the distribution than the uniform-in-component-mass model of the entire population.  

On the other hand, I might have assumed that the 8 heaviest events follow a $p(M) \propto M^{-2}$ power-law distribution between 115 \Msun$\, $ and $M_{max}^{power}$.  This might appear more consistent with the large difference between the LVK posterior on GW231123 and the next highest-mass events.   In this case, the 90\% credible interval on $M_{max}^{power}$ is $249_{-61}^{+451}$ where I take a flat prior on $M_{max}^{power}$.  The large upward shift and much greater uncertainty in the inferred $M_{max}^{power}$ illustrate the sensitivity to population model assumptions when attempting to measure the features of distributions based a limited number of events with significant measurement uncertainty. 

\section{Conclusion}

As seen in Figure \ref{fig:CDF}, the claimed mass of $238^{+28}_{-49}$ \Msun could be a very significant outlier.  It is impossible to confidently determine from the data whether this event is physically exceptional, belonging to an entirely different population; whether it is part of an underlying population distribution with very extreme tails; or whether its high mass estimate is an occasionally expected statistical artefact within a more compact population.  Bayesian model selection has been used to compare population models \citep[e.g.,][]{GWTC4:pop}, but this inevitably requires arbitrary decisions about what prior odds to assign on, say, there being one very special source coming from an entirely different astrophysical formation channel.   A variety of other techniques, such as leave-one-out methods and coarse-grained likelihoods \citep{Fishbach:2020,Essick:2022}, have been proposed to test for outliers; however, these still generally require additional assumptions to be made.

The details of the assumed population model play a key role in determining the estimated extent of the population, such as the maximum total mass of binary black holes.  If a population model is chosen and fit to the full data set, the individual events can be re-analysed with the population-inferred prior, changing their inferred properties, particularly for outlier events.  For example, for each of the population models considered here, after fitting the population model parameters, we can return to the question of the masses of GW231123; of course, its total mass will always be constrained from above to be no more than the relevant maximum total mass for a given model.  This can generally be achieved on the basis of posterior samples released by the LVK using broad priors, with search sensitivity estimated from injection analysis.

Regardless of the population model assumptions, events appearing to be most extreme are always likely to be those with the most biased posteriors.  Extra care is warranted when interpreting these extreme outliers.  As always, we must hope that future observations will allow us to determine whether apparently extreme events were statistical outliers or represented genuine population features.

\section*{Acknowledgements}

I am grateful to Arash Bahramian, Will Farr, Jonathan Gair, Yuri Levin, Eric Thrane, Alexandre Toubiana and the anonymous reviewers for very useful comments and stimulating discussions.  I acknowledge support from the Australian Research Council (ARC) Centre of Excellence for Gravitational Wave Discovery (OzGrav), through project number CE230100016. 

\bibliography{Mandel.bib}{}
\bibliographystyle{aasjournal}

\end{document}